# Performance Characterization of a Solenoid-type Gas Valve for the H$^-$ Magnetron Source at FNAL

A. Sosa[1, a)], D. S. Bollinger[1] and P. R. Karns[1]

[1]*Fermi National Accelerator Laboratory, P.O. Box 500, Batavia, IL 60510-5011, USA*

a)Corresponding author: asosa@fnal.gov

**Abstract.** The magnetron-style H$^-$ ion sources currently in operation at Fermilab use piezoelectric gas valves to function. This kind of gas valve is sensitive to small changes in ambient temperature, which affect the stability and performance of the ion source. This motivates the need to find an alternative way of feeding H$_2$ gas into the source. A solenoid-type gas valve has been characterized in a dedicated off-line test stand to assess the feasibility of its use in the operational ion sources. H$^-$ ion beams have been extracted at 35 keV using this valve. In this study, the performance of the solenoid gas valve has been characterized measuring the beam current output of the magnetron source with respect to the voltage and pulse width of the signal applied to the gas valve.

## INTRODUCTION

The H$^-$ injector at Fermi National Accelerator Laboratory (FNAL) operates continuously for up to a year or more. Requested uptime is 90% with only four weeks of planned downtime. The magnetron sources at FNAL are currently using a piezoelectric valve to pulse H$_2$ gas into the source. These valves are very reliable, however they are affected by changes in ambient temperature, which affect the amount of gas getting into the source, ultimately affecting the source's arc current and beam output. To avoid this issue, a solenoid-operated valve has been tested and characterized in an offline test stand that supports a magnetron source identical to the operational sources [1].

## PIEZOELECTRIC GAS VALVES

The gas valves currently used in the operational sources are Veeco PV-10 piezoelectric valves (Fig. 1). These valves have long been used in H$^-$ ion sources for gas injection under vacuum conditions [2, 3]. These valves start opening at 20 V, flexing proportionally to the applied voltage until they reach 100 V. However, their flexibility tends to decrease with increasing temperature. In addition, these valves show a hysteresis curve when plotting pressure vs bias voltage, as Fig. 2 shows. The typical gas pulse width used to operate the ion source is about 100 μs. The magnetron H$^-$ source at FNAL operates in pulsed mode at 15 Hz. An H$^-$ ion production cycle starts with injection of a single H$_2$ gas pulse into the ion source, as shown in Fig. 3 (a), which is electrically heated to ~200 °C and contains a cesiated cathode. Approximately 1 ms later, a 230 μs pulse provides -300 V for an arc discharge between anode and cathode, igniting a plasma. Surface produced H$^-$ ions from the cathode are extracted with a -35 kV, 150 μs pulse synchronized with the last portion of the 230 μs arc pulse.

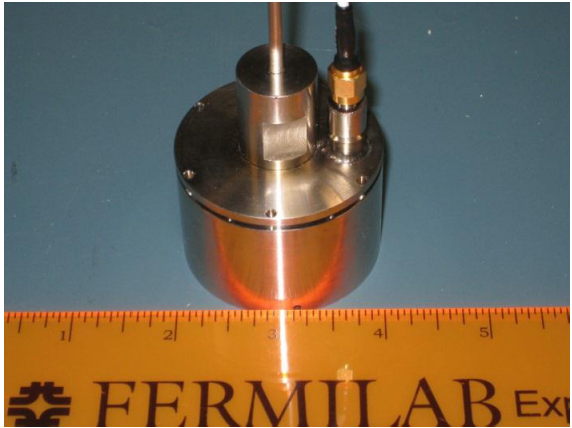
(a)
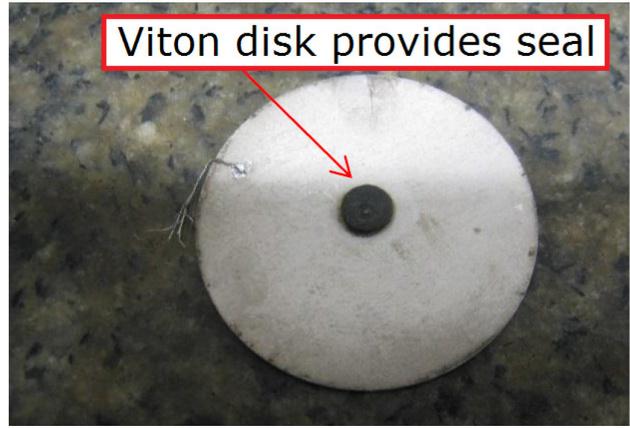
(b)

**FIGURE 1.** (a) Piezoelectric valve case. (b) Piezoelectric valve shown with the Viton disk glued on it.

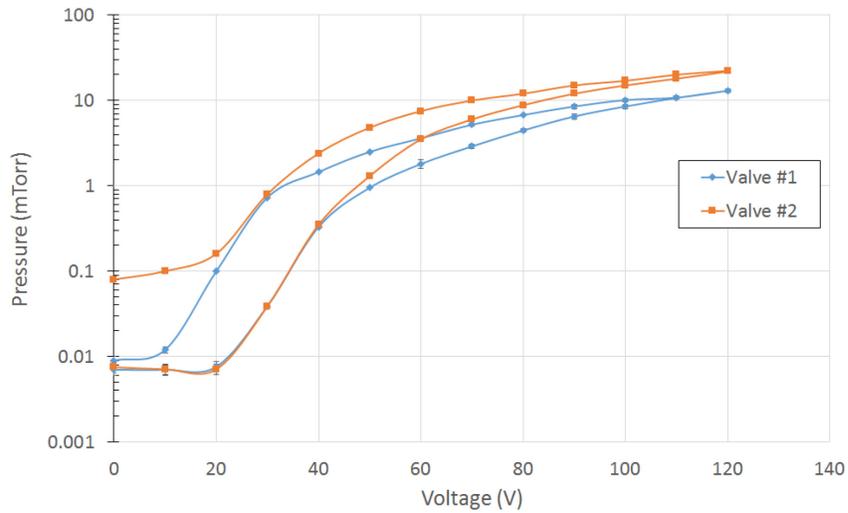

**FIGURE 2.** Hysteresis curves typically shown on two PV-10 piezoelectric valves.

In general these piezoelectric valves work well and are very reliable for our application, but their flexibility changes with temperature. As room temperature increases, the piezoelectric valve flexes less, allowing less gas into the source, thus affecting the arc current and beam output. This means a 1 °C change in room temperature can translate into a 1 µTorr change in vacuum pressure and as much as 1.5 A change in arc current. This inconvenience motivated the search for an alternative gas valve able to pulse gas into the source and avoid its temperature dependence.

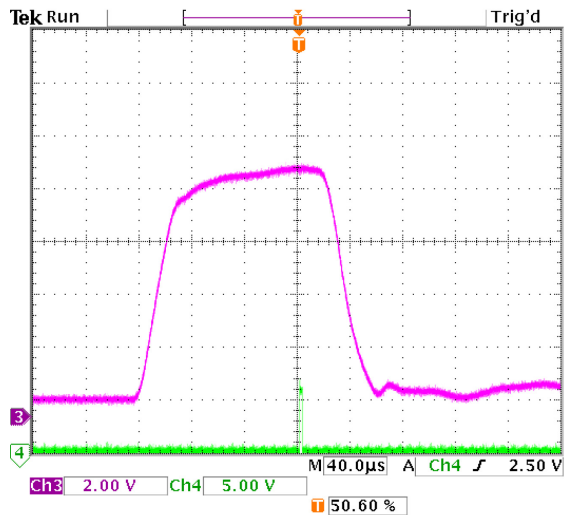 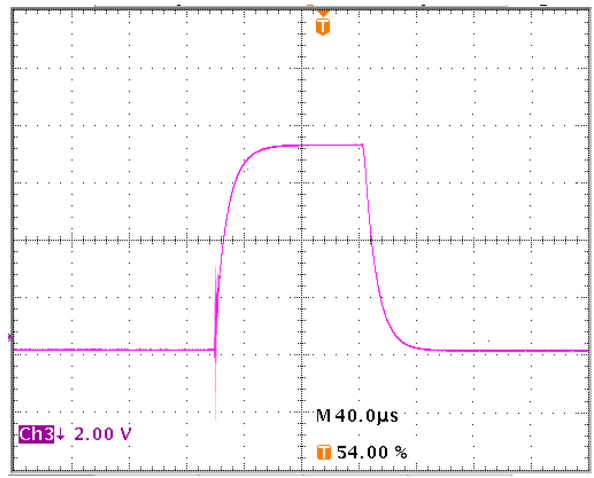

(a)                                      (b)

**FIGURE 3.** (a) Scope trace of the $H_2$ gas pulse (magenta) using a piezoelectric gas valve. Scale is 20 V/div. (b) Scope trace of the $H_2$ gas pulse (magenta) using a solenoid gas valve. Scale is 60 V/div.

## SOLENOID GAS VALVES

A pulse valve Series 9 from Parker (Fig. 4) has been installed in a magnetron $H^-$ ion source in an offline test stand. A VESPEL poppet was used due to its thermal and mechanical properties. The valve aperture is 0.5 mm in diameter. Preliminary beam runs allowed the source to operate in similar conditions as when using piezoelectric valves. The typical gas pulse width used with the piezoelectric valves is about 90 to 115 µs, compared to a range of 100 to 150 µs when using the solenoid gas valve in order to obtain similar arc current and beam output. When running the ion source with this valve, biased at 275V and 124 µs gas pulse width set 840 µs before the arc discharge, the beam output was maximum. Vacuum pressure, extraction voltage, arc voltage and arc current remained constant during this test at 4.2 µTorr, -35 kV, 189 V and 15 A, respectively. In these conditions, the arc pulse width was increased from 100 to 275 µs and beam current and cathode temperature were recorded. Results are plotted in Fig. 5. An arc pulse width ranging from 150 to 250 µs yielded the highest beam currents, ceteris paribus. In this range the ion source can deliver beam currents above 50 mA, which satisfies the needs of the injection line downstream.

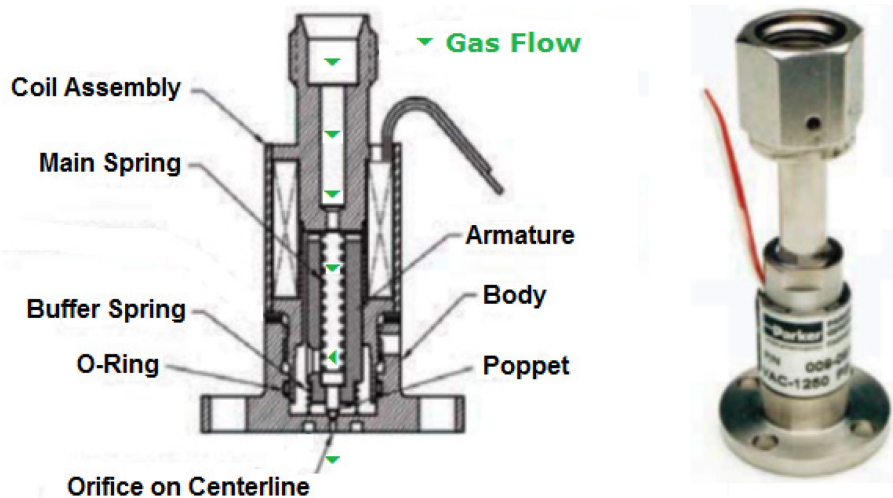

**FIGURE 4.** Solenoid gas valve used for tests with $H^-$ source.

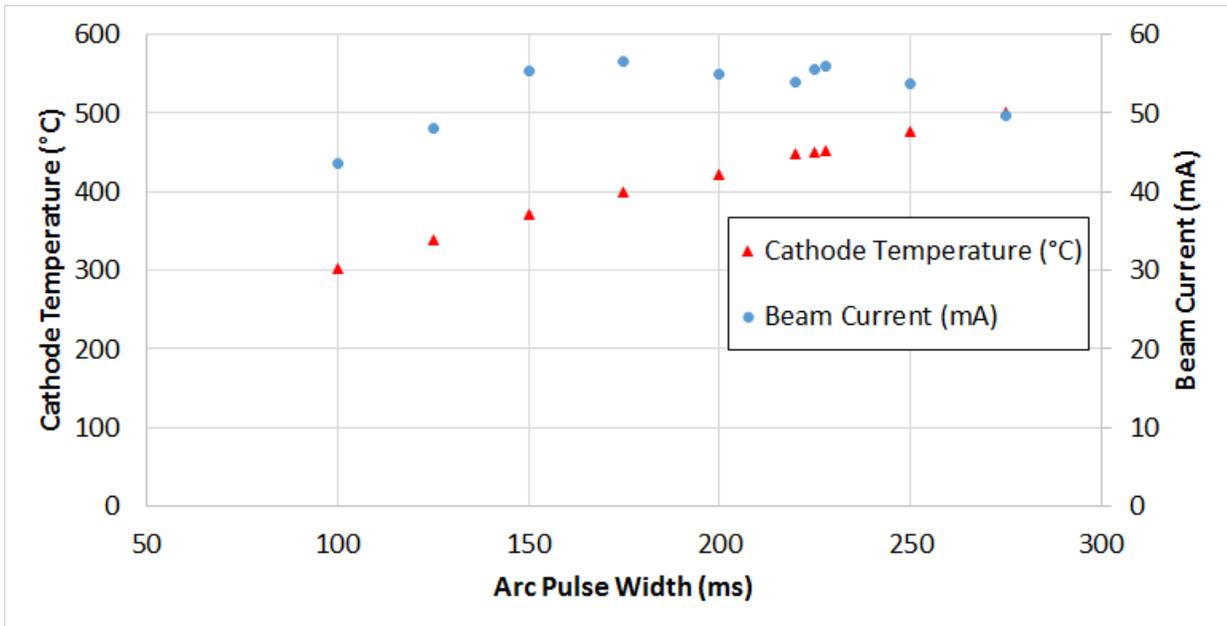

**FIGURE 5.** Beam current and cathode temperature vs arc pulse width using the solenoid gas valve in the H⁻ source.

## CONCLUSIONS

The Pulse Valve Series 9 from Parker has been successfully integrated into a magnetron H⁻ ion source. This type of valve lacks the hysteresis cycle and temperature dependence of the piezoelectric valve. Its performance is comparable to the piezoelectric valve currently used, sustaining beam currents above 50 mA and satisfying the needs of the operational ion sources at FNAL.

## ACKNOWLEDGMENTS

This research was supported by Fermi Research Alliance, LLC under Contract No. De-AC02-07CH11359 with the United States Department of Energy. We would like to thank Andrew Feld and Kenneth Koch for their technical support with the gas valves and the ion source during this experiment.